\documentclass[twocolumn,fp,a4paper]{jpsj3}

\usepackage{subfigure}
\usepackage{wrapfig}
\usepackage{color}
\usepackage{graphicx}
\usepackage{dcolumn}
\usepackage{bm}
\usepackage{txfonts}

\def\journal #1#2#3#4{#1 {\bf #2}, #3 (#4)}

\title{Weyl Semimetal in the Strong Coulomb Interaction Limit}

\author{\name{Akihiko \surname{Sekine}}\thanks{E-mail: sekine@imr.tohoku.ac.jp} and \name{Kentaro \surname{Nomura}}}
\inst{Institute for Materials Research, Tohoku University, Sendai 980-8577, Japan} 

\abst{We study the effects of strong $1/r$ long-range Coulomb interactions in a Weyl semimetal.
We consider a three-dimensional (3D) Dirac fermion system on a lattice with a time-reversal symmetry breaking term, and take into account $1/r$ long-range Coulomb interactions between the bulk electrons.
This model is regarded as the case where magnetic impurities are doped into the bulk of a 3D topological insulator.
With the use of the strong coupling expansion of the lattice gauge theory and the mean-field approximation, we analyze the system from the strong coupling limit.
It is shown that parity symmetry of the system is spontaneously broken in the strong coupling limit, and a different type of the Weyl semimetal, in which time-reversal and parity symmetries are broken, appears in the strong coupling limit.
A possible global phase diagram of a correlated Weyl semimetal is presented.}

\begin{document}
\maketitle

\section{Introduction}
As a novel topological phase of matter, the Weyl semimetal has attracted much attention since a theoretical prediction was made.\cite{Wan2011}
Weyl semimetals have three-dimensional (3D) linear dispersions near the band touching points, the Weyl nodes.
These linear dispersions are described by the Weyl fermions, namely by the Weyl Hamiltonian $\mathcal{H}(\bm{k})=\sum_{i=1}^3\bm{v}_i\cdot\bm{k}\sigma_i$ where $\sigma_i$ are the Pauli matrices.
The chirality is defined by $c={\rm sgn}[\bm{v}_1\cdot(\bm{v}_2\times\bm{v}_3)]=\pm1$.
Breaking of either time-reversal or inversion symmetry is required to realize the Weyl semimetal.\cite{Volovik2003,Volovik2007,Murakami2007,Burkov2011a,Burkov2011,Halasz2012,Zyuzin2012a,Lin2013}
Since all the three Pauli matrices are used, the Weyl fermion cannot be massive by itself and the existence of a single Weyl node is robust against perturbations.
In inversion symmetric systems, two Weyl nodes with opposite chirality exist at the momentum points $\pm\bm{k}_0$.
The Weyl nodes disappear and the energy gap opens only when the two Weyl nodes with opposite chirality meet each other.
Such a topological feature is characterized by the surface states\cite{Wan2011,Burkov2011a,Delplace2012}
and the magnetoelectric response which is described by the $\theta$ term.\cite{Zyuzin2012,Wang2013,Vazifeh2013}
Other properties such as the anomalous Hall effect,\cite{Burkov2011a} quantum Hall effect,\cite{Yang2011} charge transport,\cite{Hosur2012,Ashby2013,Rosenstein2013} superconductivity,\cite{Meng2012,Cho2012} and correlation effects\cite{Wei2012,Sekine2013,Jho2013} have also been studied.

Since electron correlation has been revealed to be important in many systems and strongly correlated systems have been one of the central research subjects in condensed matter physics,
it is notable that the Weyl semimetal phases are predicted in pyrochlore iridates, strongly correlated 5$d$-electron systems with spin-orbit coupling.\cite{Wan2011,Witczak-Krempa2012}
It is natural to expect that electron correlation plays important roles to realize topologically nontrivial phases.
Actually there are many theoretical studies which show that topologically nontrivial phases are induced by the intermediate strength of short-range correlation.\cite{Raghu2008,Sun2009,Wen2010,Kurita2011,Ruegg2011,Yang2011a}
However, these phases are broken to be topologically trivial by sufficiently strong short-range correlation.\cite{Wan2011,Witczak-Krempa2012,Raghu2008,Sun2009,Wen2010,Kurita2011,Ruegg2011,Yang2011a}

As for the effects of short-range correlation in topological insulators, many studies show that topological insulator phases are not stable against strong short-range correlation.\cite{Pesin2010,Rachel2010,Varney2010,Hohenadler2011,Yu2011,Yamaji2011,Zheng2011,Hohenadler2012,Yoshida2012}
On the other hand, as for long-range correlation, recent studies show that topological insulator phases survive even in the limit of strong $1/r$ long-range correlation.\cite{Sekine2013a,Araki2013}
Are Weyl semimetals stable against long-range correlation?
Motivated by this question, we undertook the present work.

Electron correlation effects in graphene, a two-dimensional Dirac fermion system, have been studied intensively.\cite{Khveshchenko2001,Gorbar2002,Khveshchenko2004,Son2007,Hands2008,Khveshchenko2009,Drut2009,Drut2009a,Armour2010,Araki2010,Armour2011,Giedt2011,Wang2012,Giuliani2012,Buividovich2012}
When the system is described by the quantum electrodynamics, due to the smallness of the Fermi velocity, Dirac fermions interact only with the electric field, i.e., interact via $1/r$ Coulomb interactions.
In this case, the effective coupling becomes strong, and the strength is determined by the dielectric constant of the substrate.
By considering the system on a lattice, the strong coupling lattice gauge theory can be applied.\cite{Drut2009,Drut2009a,Armour2010,Araki2010,Armour2011,Giedt2011,Buividovich2012}
Studies based on the lattice gauge theory show that at sufficiently strong coupling, the system becomes insulating due to spontaneous chiral (sublattice) symmetry breaking.
An analytical result for the value of the mass gap in the strong coupling limit\cite{Araki2010} is in good agreement with Monte Carlo results.\cite{Drut2009,Drut2009a,Armour2010}

The purpose of this study is (1) to reveal whether the Weyl semimetal with two Weyl nodes, the minimal number of the nodes in the lattice models, is stable against strong $1/r$ long-range Coulomb interactions,
and (2) to present a possible global phase diagram of a correlated Weyl semimetal.

In this paper, we study the effects of strong electron correlation in a Weyl semimetal where the Hamiltonian is described by the four-component Dirac fermions with a time-reversal symmetry breaking perturbation term.
We adopt $1/r$ long-range Coulomb interactions as the interactions between the bulk electrons, because the screening effect is considered to be weak near the Fermi level in Dirac fermion systems due to the vanishing density of states. 
Based on the U(1) lattice gauge theory, we perform the strong coupling expansion, and analyze the system from the strong coupling limit with the mean-field approximation.

This paper is organized as follows.
In Sec. \ref{Sec-Model}, we formulate a model of an interacting Weyl semimetal according to the quantum electrodynamics.
In Sec. \ref{Sec-SCE}, we derive the effective action in the strong coupling limit.
In Sec. \ref{Sec-FE}, we present all the possible instabilities within the mean-field approximation.
Then we derive the free energies at zero temperature with the use of the Hubbard-Stratonovich transformation.
In Sec. \ref{Sec-Results}, we show the numerical results.
The ground state is determined by minimizing the free energies.
In Sec. \ref{Sec-Phasediagram}, we consider a possible global phase diagram of our model.
In Sec. \ref{Sec-Discussion} we discuss our results and finally in Sec. \ref{Sec-Summary}, we summarize this study.

\section{Model \label{Sec-Model}}
Let us start from a continuum model for a non-interacting Weyl semimetal.
We consider a Weyl semimetal which consists of two Weyl nodes separated in the momentum space.
In this case, the general Hamiltonian is written by the 3D Dirac fermion,
\begin{align}
\begin{split}
\mathcal{H}_0(\bm{k})=v_{\rm F}\bm{k}\cdot\bm{\alpha}+m_0\alpha_4+\bm{b}\cdot\bm{\Sigma},\label{H-conti}
\end{split}
\end{align}
where $v_{\rm F}$ is the Fermi velocity, $m_0$ is the mass of the Dirac fermion, and $\alpha_j$ $(j=1,2,3,4)$ are the $4\times4$ alpha matrices.
The term $\bm{b}\cdot\bm{\Sigma}$, with $\bm{\Sigma}=-i/2([\alpha_2,\alpha_3],[\alpha_3,\alpha_1],[\alpha_1,\alpha_2])$, is a time-reversal symmetry breaking perturbation.
Without loss of generality, we can set $\bm{b}=(0,0,b)$.
The eigenvalues of this Hamiltonian are obtained analytically, $E(\bm{k})=\pm\sqrt{v_{\rm F}^2(k_1^2+k_2^2)+(\sqrt{v_{\rm F}^2k_3^2+m_0^2}\pm b)^2}$,
and we see that the band touching points (the Weyl nodes) $W_\pm$ are given by $W_\pm=(0,0,\pm\sqrt{b^2-m_0^2})$ with $b>|m_0|$.
Then the energy bands near the Weyl nodes are found to be $E(\bm{q})=\pm v_{\rm F}\sqrt{q_1^2+q_2^2+(1-m_0^2/b^2)q_3^2}$ where $\bm{q}=\bm{k}-W_\pm$.
This is just the dispersion of the Weyl fermions, since this can be obtained by diagonalizing the Weyl Hamiltonian $\mathcal{H}(\bm{q})=\sum_{i=1}^3\bm{v}_i\cdot\bm{q}\sigma_i$.

Next let us consider the Hamiltonian (\ref{H-conti}) on a cubic lattice.
In the following, as the unperturbed part, we adopt the Wilson fermion which is known as a lattice model without fermion doublers.
The Wilson fermion can also be regarded as the effective lattice model for 3D topological insulators such as Bi$_2$Se$_3$.\cite{Zhang2009,Liu2010}
Hence, our model is regarded as the case where magnetic impurities are doped into the bulk of a 3D topological insulator, because the term $b\Sigma_3$ can be considered as the exchange (ferromagnetic) coupling between the bulk electrons and magnetic impurities.
The non-interacting Hamiltonian we consider is given by
\begin{align}
\begin{split}
\mathcal{H}_0(\bm{k})=v_{\rm F}\sum_{j=1}^3\alpha_j\sin k_j+m(\bm{k})\alpha_4+b\Sigma_3,\label{H0-lattice}
\end{split}
\end{align}
where $m(\bm{k})=m_0+r\sum_{j=1}^3(1-\cos k_j)$.
The alpha matrices are given by the Dirac representation, $\alpha_j=\tau_1\otimes \sigma_j$ and $\alpha_4=\tau_3\otimes \bm{1}$ where the Pauli matrices $\tau_j$ and $\sigma_j$ denote the orbital and spin degree of freedom, respectively.
Namely, the spinor in the Hamiltonian (\ref{H0-lattice}) is written in the basis of $\psi_{\bm k}^\dag=[c^\dag_{\bm{k}+\uparrow},c^\dag_{\bm{k}+\downarrow},c^\dag_{\bm{k}-\uparrow},c^\dag_{\bm{k}-\downarrow}]$, where $c^\dag$ is the creation operator of an electron, $+,-$ denote two orbitals, and $\uparrow$ ($\downarrow$) denotes up-(down-)spin.
The matrix $\Sigma_3$ is given explicitly as $\Sigma_3=\bm{1}\otimes\sigma_3$.
The Weyl nodes appear where the wave vector $\bm{k}$ satisfies the condition $b^2=[m_0+r{\sum}_{j}(1-\cos k_j)]^2+\sin^2 k_3$ and $\sin k_1=\sin k_2=0$.
$k_1$ and $k_2$ can take the value 0 or $\pi$.
Setting $|m_0|$ small, we concentrate on the Weyl nodes which appear on the $(k_1,k_2)=(0,0)$ line.

We introduce $1/r$ Coulomb interactions between the bulk electrons according to the U(1) lattice gauge theory (lattice quantum electrodynamics).
In condensed matter, the Fermi velocity is rather small compared to the speed of light $c$, i.e., $v_{\rm F}/c\sim 10^{-3}$.
In this case, the interactions with the spatial components of the four-vector potential are suppressed by the factor $v_{\rm F}/c$, and thus we can regard that the Dirac fermions interact only via the electric field i.e. $1/r$ Coulomb interactions.
Then the Euclidean action of the system is given by
\begin{align}
\begin{split}
S&=S_F^{(\tau)}+S_F^{(s)}+(m_0+3r+r_\tau)\sum_{n}\bar{\psi}_n \psi_n\\
&\quad+b\sum_n\bar{\psi}_n\gamma_0\Sigma_3 \psi_n+S_G,\label{Action}
\end{split}
\end{align}
where $S_F^{(\tau)}+S_F^{(s)}$ is the fermionic part without the mass term,
\begin{align}
\begin{split}
S_F^{(\tau)}&=-\sum_{n}\left[\bar{\psi}_nP^-_0 U_{n,0}\psi_{n+\hat{0}} + \bar{\psi}_{n+\hat{0}}P^+_0 U^\dag_{n,0}\psi_n\right],\\
S_F^{(s)}&=-\sum_{n,j}\left[\bar{\psi}_nP^-_j\psi_{n+\hat{j}} + \bar{\psi}_{n+\hat{j}}P^+_j\psi_n\right],
\end{split}
\end{align}
and $S_G$ is the pure U(1) gauge part,
\begin{align}
\begin{split}
S_G=\beta\sum_n\sum_{\mu>\nu}\left[1-\frac{1}{2}\left(U_{n,\mu}U_{n+\hat{\mu},\nu}U^\dag_{n+\hat{\nu},\mu}U^\dag_{n,\nu}+{\rm H.c.}\right)\right].
\end{split}
\end{align}
Here $\bar{\psi}=\psi^\dag\gamma_0$, $n=(n_0,n_1,n_2,n_3)$ denotes a spacetime lattice site, $\hat{\mu}$ denotes the unit vector along the $\mu$ direction, and $P^\pm_\mu=(r_\mu\pm\gamma_\mu)/2$ with $r_0=r_\tau$ and $r_1=r_2=r_3=r$.
$U_{n,\mu}$ are the U(1) link variables with $U_{n,j}=1$ and $U_{n,0}=e^{i\theta_n}\ (-\pi\leq\theta_n\leq\pi)$.
The timelike Wilson term (the term proportional to $r_\tau$) is introduced to the unperturbed part (the $b=0$ part) to eliminate the fermion doublers.
This is because when $b=0$, the system should possess a single Dirac cone around the $\Gamma$ point.
According to the non-interacting Hamiltonian of 3D topological insulators such as Bi$_2$Se$_3$, the gamma matrices $\gamma_\mu$ are given by the Dirac representation.

A parameter $\beta$ which represents the strength of $1/r$ Coulomb interactions is given by
\begin{align}
\begin{split}
\beta=\frac{v_{\rm F}\epsilon}{e^2}=\frac{v_{\rm F}\epsilon}{4\pi c\alpha},
\end{split}
\end{align}
where $\epsilon$ is the dielectric constant of the system, $e$ is the electric charge, and $\alpha(\simeq 1/137)$ is the fine-structure constant.
In the following, we consider the case with $\beta\ll 1$ i.e., the case with small dielectric constant.

\section{Strong Coupling Expansion \label{Sec-SCE}}
Let us analyze the system from the strong coupling limit ($\beta=0$).
We derive the effective action which is defined by carrying out the integration with respect to the gauge field variables $U_{n,0}$:
\begin{align}
\begin{split}
Z=\int \mathcal{D}[\psi,\bar{\psi},U_0]e^{-S}=\int \mathcal{D}[\psi,\bar{\psi}]e^{-S_{\rm eff}}.\label{partition-function}
\end{split}
\end{align}
In the strong coupling limit, $U_{n,0}$ is contained only in $S_F^{(\tau)}$.
Then the integral $\int \mathcal{D}U_0e^{-S_F^{(\tau)}}$ is evaluated as
\begin{align}
\begin{split}
&\prod_n\int_{-\pi}^\pi\frac{d\theta_n}{2\pi}\exp\left[{\bar{\psi}_nP^-_0 U_{n,0}\psi_{n+\hat{0}} + \bar{\psi}_{n+\hat{0}}P^+_0 U^\dag_{n,0}\psi_n}\right]\\
&=\prod_n\left[1+\bar{\psi}_nP^-_0 \psi_{n+\hat{0}}\bar{\psi}_{n+\hat{0}}P^+_0 \psi_n+\cdots\right]\\
&\approx e^{\sum_n\bar{\psi}_nP^-_0 \psi_{n+\hat{0}}\bar{\psi}_{n+\hat{0}}P^+_0 \psi_n},
\end{split}
\end{align}
where we have used the property of the Grassmann variables $\psi_\alpha$ and $\bar{\psi}_\alpha$, $\psi_\alpha^2=\bar{\psi}_\alpha^2=0$ with $\alpha$ denoting the component of the spinors.
In the second line, we have neglected the terms which consist of 8, 12, and 16 different Grassmann variables.
The contributions of those terms appear in higher orders of the order parameters, and thus the results will not be changed qualitatively even if those terms are taken into account.
We can rewrite the exponent as
\begin{align}
\begin{split}
\bar{\psi}_{n,\alpha}(P^-_0)_{\alpha\beta} \psi_{n+\hat{0},\beta}\bar{\psi}_{n+\hat{0},\gamma}(P^+_0)_{\gamma\delta} \psi_{n,\delta}\\
= -\mathrm{tr}\left[N_nP^+_0N_{n+\hat{0}}P^-_0\right],
\end{split}
\end{align}
where we have defined $(N_n)_{\alpha\beta}=\bar{\psi}_{n,\alpha}\psi_{n,\beta}$ and used $(P^\pm_0)_{\alpha\beta}=(P^\pm_0)_{\beta\alpha}$. The subscripts $\alpha$ and $\beta$ denote the component of the spinors.
In general, we can perform the integration with respect to the SU($N$) gauge field variables $U$ in Eq. (\ref{partition-function}) by using the SU($N$) group integral formulas: $\int dU 1=1$, $\int dU U_{ab}=0$, $\int dU U_{ab}U_{cd}^\dagger=\delta_{ad}\delta_{bc}/N$, and so on.
We obtain the effective action in the strong coupling limit given by
\begin{align}
\begin{split}
S_{\mathrm{eff}}&=-\sum_{n,j}\left[\bar{\psi}_nP^-_j\psi_{n+\hat{j}} + \bar{\psi}_{n+\hat{j}}P^+_j\psi_n\right]\\
&\quad+\sum_n\mathrm{tr}\left[N_nP^+_0N_{n+\hat{0}}P^-_0\right]\\
&\quad+(m_0+3r+r_\tau)\sum_{n}\bar{\psi}_n\psi_n+b\sum_n\bar{\psi}_n\gamma_0\Sigma_3 \psi_n.\label{Effective-action}
\end{split}
\end{align}
From this equation, we see that the electron-electron interactions in the strong coupling limit are spatially on-site interactions but not in (imaginary) time.
As we shall see below, the competition between this effective on-site interactions and the exchange interactions occurs.

\section{Possible Instabilities and the Free Energies in the Strong Coupling Limit \label{Sec-FE}}
We apply the extended Hubbard-Stratonovich transformation to derive the free energy.
Introducing the two complex auxiliary fields $Q$ and $Q'$ (these are matrices), $e^{\kappa\mathrm{tr}AB}$ with $\kappa>0$ and $A,B$ being matrices is deformed as follows:\cite{Sekine2013a}
\begin{align}
\begin{split}
&e^{\kappa\mathrm{tr}AB}={\rm (const.)}\times\\
&\int \mathcal{D}[Q,Q'] \exp\left\{-\kappa\left[Q_{\alpha\beta}Q'_{\alpha\beta}-A_{\alpha\beta}Q_{\beta\alpha}-B^T_{\alpha\beta}Q'_{\beta\alpha}\right]\right\},\label{EHS}
\end{split}
\end{align}
where the superscript $T$ denotes the transpose of a matrix.
This integral is approximated by the saddle point values $Q_{\alpha\beta}=\left\langle B^T\right\rangle_{\beta\alpha}$ and $Q'_{\alpha\beta}=\left\langle A\right\rangle_{\beta\alpha}$.

We set $(\kappa, A, B)=(1, N_nP^+_0, -N_{n+\hat{0}}P^-_0)$ to decouple the interaction term (the second term) in Eq. (\ref{Effective-action}) to fermion bilinear form.
In this process, we assume the matrix form of $\langle N_n\rangle$ with the mean-field approximation.
Let us recall that, in the formalism of the quantum electrodynamics, the form is restricted to the sum of independent 16 matrices which are consist of the $4\times 4$ gamma matrices.
The 16 matrices are given as follows: (i) scalar $\bm{1}$, (ii) vector $\gamma_\mu$ ($\mu=0,1,2,3$), (iii) tensor $\sigma_{\mu\nu}=\frac{i}{2}[\gamma_\mu,\gamma_\nu]$, (iv) pseudovector $\gamma_\mu\gamma_5$, (v) pseudoscalar $\gamma_5$, where these terminology comes from how the terms are transformed under the Lorentz transformation.

\subsection{Fermion bilinears and instabilities}
Here we give the explicit expression of the possible instabilities.
As shown later, the interaction term in the effective action (\ref{Effective-action}) is written like $\bar{\psi}_n\langle N_n\rangle\psi_n$ after the Hubbard-Stratonovich transformation.
Then, to be specific and for simplicity, let us consider the Hamiltonian of a continuum model
\begin{align}
\begin{split}
\mathcal{H}(\bm{k})=\mathcal{H}_0(\bm{k})+\gamma_0\langle N_n\rangle
\end{split}
\end{align}
with $\mathcal{H}_0(\bm{k})=k_j\alpha_j+m_0\alpha_4+b\Sigma_3$ [Eq. (\ref{H-conti})].
Note that the following terminology of the instabilities is not for $\gamma_0\langle N_n\rangle$, but for $\langle N_n\rangle$.
\\$\bullet$ {\it Scalar component:}
$\gamma_0\langle N_n\rangle=\sigma\alpha_4$. Here we have used the fact that $\gamma_0=\alpha_4$. In this case, it is found that the bare mass $m_0$ is renormalized to be $m_0+\sigma$.
\\$\bullet$ {\it Vector component:}
$\gamma_0\langle N_n\rangle=p_\mu\alpha_\mu$ ($\alpha_0\equiv \bm{1}$). Here we have used the fact that $\gamma_0\gamma_j=\alpha_j$ and $\gamma_0^2=\bm{1}$. In this case, it is found that the wave vectors $k_j$ are shifted to be $k_j+p_j$, and that the energy level is shifted by $p_0$.
\\{$\bullet$ \it Tensor component:}
$\gamma_0\langle N_n\rangle=\frac{i}{2}c_{\mu\nu}\gamma_0[\gamma_\mu,\gamma_\nu]$. The six matrices $\gamma_0\langle N_n\rangle$ are given explicitly by the Dirac representation as
\begin{align}
\begin{split}
\Sigma'_j\equiv
\begin{bmatrix}
\sigma_j & 0\\
0 & -\sigma_j
\end{bmatrix},\ \ \ \ \
\Pi_j\equiv i
\begin{bmatrix}
0 & \sigma_j\\
-\sigma_j & 0
\end{bmatrix}.
\end{split}
\end{align}
\\$\bullet$ {\it Pseudovector component:}
$\gamma_0\langle N_n\rangle=d_\mu\gamma_0\gamma_\mu\gamma_5$. The four matrices $\gamma_0\langle N_n\rangle$ are given explicitly by the Dirac representation as
\begin{align}
\begin{split}
\Sigma_j\equiv
\begin{bmatrix}
\sigma_j & 0\\
0 & \sigma_j
\end{bmatrix},\ \ \ \ \
\Pi_0\equiv
\begin{bmatrix}
0 & 1\\
1 & 0
\end{bmatrix}.
\end{split}
\end{align}
\\$\bullet$ {\it Pseudoscalar component:}
$\gamma_0\langle N_n\rangle=\Delta\alpha_5$. The matrix $\alpha_5$($\equiv -i\gamma_0\gamma_5$) is given explicitly by the Dirac representation as $\alpha_5=\tau_2\otimes\bm{1}$, and this anticommutes with the other four alpha matrices, i.e. the Clifford algebra $\{\alpha_\mu,\alpha_\nu\}=2\delta_{\mu\nu}$ is satisfied.

Let us look at the properties of the matrices $\Sigma_j$, $\Sigma'_j$, $\Pi_\mu$, and $\alpha_5$.
In the Dirac representation, the time-reversal operator $\mathcal{T}$ and the parity (spatial inversion) operator $\mathcal{P}$ are given by $\mathcal{T}=\bm{1}\otimes(-i\sigma_2)\mathcal{K}$ and $\mathcal{P}=\sigma_3\otimes\bm{1}$, where $\mathcal{K}$ is the complex conjugation operator.
It is easily shown that $\Sigma_j$ and $\Sigma'_j$ are odd under time-reversal but even under parity:
\begin{align}
\begin{split}
\mathcal{T}\Sigma_j\mathcal{T}^{-1}=\mathcal{T}\Sigma'_j\mathcal{T}^{-1}=-1,\ \  \mathcal{P}\Sigma_j\mathcal{P}^{-1}=\mathcal{P}\Sigma'_j\mathcal{P}^{-1}=+1.
\end{split}
\end{align}
On the other hand, $\Pi_\mu$ are even under time-reversal but odd under parity:
\begin{align}
\begin{split}
\mathcal{T}\Pi_\mu\mathcal{T}^{-1}=+1,\ \ \ \mathcal{P}\Pi_\mu\mathcal{P}^{-1}=-1.
\end{split}
\end{align}
$\alpha_5$ is odd under both time-reversal and parity:
\begin{align}
\begin{split}
\mathcal{T}\alpha_5\mathcal{T}^{-1}=-1,\ \ \ \mathcal{P}\alpha_5\mathcal{P}^{-1}=-1.
\end{split}
\end{align}
In our model, the spinor is written in the basis of $\psi_{\bm k}^\dag=[c^\dag_{\bm{k}+\uparrow},c^\dag_{\bm{k}+\downarrow},c^\dag_{\bm{k}-\uparrow},c^\dag_{\bm{k}-\downarrow}]$
where $+,-$ denote the two orbitals and $\uparrow$ ($\downarrow$) denotes up(down)-spin.
Then we see that $\Sigma_j$ represents the ferromagnetism, and that $\Sigma'_j$ represents a kind of the ``antiferromagnetism''.
As for $\Pi_\mu$ and $\alpha_5$, the physical interpretation is somewhat difficult.
We see that $\Pi_0$ and $\alpha_5$ represent spin-independent orbital-ordered states, and $\Pi_j$ represents a spin-dependent orbital-ordered state.

To study the stability of the Weyl semimetals, let us consider which matrices we should take into account among these 16 matrices as a mean-field ansatz for $\langle N_n\rangle$.
First of all, the mass renormalization, i.e. the identity matrix $\bm{1}$ term must be considered.
The $\gamma_\mu$ terms, which result in the momentum and energy level shifts, could be dropped.
In addition, note that the non-interacting Hamiltonian (\ref{H0-lattice}) contains the $\Sigma_3$ term.
This indicates that the $\gamma_0\Sigma_j$ and $\gamma_0\Sigma_j'$ terms can also be dropped.
Furthermore, it is known that these terms can generate other Weyl semimetal phases rather than open band gaps.\cite{Burkov2011}
On the other hand, the $\gamma_0\Pi_\mu$ and $i\gamma_5$ terms can break the Weyl semimetal phase and open band gaps.
In the following, we take into account the $\bm{1}$, $\gamma_0\Pi_\mu$, and $i\gamma_5$ terms as the matrix form of $\langle N_n\rangle$.

\subsection{The $\alpha_5$ instability}
In this case, we assume that $\langle N_n\rangle=-\sigma \bm{1}+\Delta i\gamma_5$.
The interaction term is decoupled according to Eq. (\ref{EHS}).
After a calculation, we obtain
\begin{align}
\begin{split}
&e^{-\sum_n\mathrm{tr}\left[N_nP^+_0N_{n+\hat{0}}P^-_0\right]}\\
&\sim\exp\left\{-\sum_{n}\left[(1-r_\tau^2)\sigma^2+(1+r_\tau^2)\Delta^2+\bar{\psi}_n\Gamma\psi_n\right]\right\},\label{int-term}
\end{split}
\end{align}
where $\Gamma=\frac{1}{2}\left[(1-r_\tau^2)\sigma+i\gamma_5^T(1+r_\tau^2)\Delta\right]$.
We are now in a position to derive the free energy at zero temperature in the strong coupling limit.
Combining Eqs. (\ref{Effective-action}) and (\ref{int-term}), the effective action expressed by the auxiliary fields $\sigma$ and $\Delta$ is given by
\begin{align}
\begin{split}
S_{\mathrm{eff}}(\sigma,\Delta)&={\textstyle \frac{V}{T}}\left[(1-r_\tau^2)\sigma^2+(1+r_\tau^2)\Delta^2\right]\\
&\quad+\sum_k \bar{\psi}_k\mathcal{M}(\bm{k};\sigma,\Delta)\psi_k, \label{S_eff_SCL}
\end{split}
\end{align}
where the matrix $\mathcal{M}$ is given explicitly as
\begin{align}
\begin{split}
\mathcal{M}&=
\begin{bmatrix}
\tilde{m}(\bm{k})+r_\tau+b\sigma_3 & \sigma_j\sin k_j+i\tilde{\Delta}\\
-\sigma_j\sin k_j+i\tilde{\Delta} & \tilde{m}(\bm{k})+r_\tau-b\sigma_3
\end{bmatrix}
\equiv
\begin{bmatrix}
A & B\\
C & D
\end{bmatrix}.\label{M}
\end{split}
\end{align}
Here $\tilde{\Delta}=\frac{1}{2}(1+r_\tau^2)\Delta$, $V$ and $T$ are the volume and temperature of the system, respectively, and we have done the Fourier transform from $n=(n_0,\bm{n})$ to $k=(k_0,\bm{k})$.
The term $\tilde{m}(\bm{k})$ is given by
\begin{align}
\begin{split}
\tilde{m}(\bm{k})=m_0+{\textstyle \frac{1}{2}}(1-r_\tau^2)\sigma+r\sum_{j}\left(1-\cos k_j\right). \label{m_k-tilde}
\end{split}
\end{align}
This term is understood as a term to which $m(\bm{k})$ in the non-interacting Hamiltonian (\ref{H0-lattice}) changes in the strong coupling limit.
$r_\tau$ in $A$ and $D$ of Eq. (\ref{M}) comes from the timelike Wilson term of the original action.
From the effective action, we derive the free energy at zero temperature per unit of spacetime volume, according to the usual formula $\mathcal{F}=-\frac{T}{V}\ln Z$.
The partition function $Z$ is calculated by the Grassmann integral formula $Z=\int D[\psi,\bar{\psi}]e^{-\bar{\psi}\mathcal{M}\psi}=\mathrm{det}\mathcal{M}$ and the determinant of $\mathcal{M}$ is calculated by the formula $\mathrm{det}\mathcal{M}=\mathrm{det}A\cdot\mathrm{det}\left(D-CA^{-1}B\right)$.
After a straightforward calculation, finally we arrive at the free energy in the strong coupling limit:
\begin{align}
\begin{split}
&\mathcal{F}(\sigma,\Delta)=(1-r_\tau^2)\sigma^2+(1+r_\tau^2)\Delta^2-\int_{-\pi}^{\pi}\frac{d^3k}{(2\pi)^3}\\
&\times\ln\left\{\left[s^2(\bm{k})+\left[\tilde{m}(\bm{k})+r_\tau\right]^2+\tilde{\Delta}^2-b^2\right]^2+4b^2s^2_\bot(\bm{k})\right\},\label{Feff-SCL}
\end{split}
\end{align}
where $s^2(\bm{k})=\sum_{j=1}^3\sin^2k_j$, and $s^2_\bot(\bm{k})=\sum_{l=1}^2\sin^2k_l$.
The ground state is determined by the stationary condition $\partial \mathcal{F}(\sigma,\Delta)/\partial \sigma=\partial \mathcal{F}(\sigma,\Delta)/\partial \Delta=0$.

Note that the free energy at $b=0$ corresponds to that of the Wilson fermions with $1/r$ Coulomb interactions in the strong coupling limit.\cite{Sekine2013a}
When $r_\tau>1$, the free energy doesn't have the stationary point, because both the first and third term decrease the value of $\mathcal{F}(\sigma, \Delta)$ with increasing $\sigma$.
This phenomenon is known as the reflection positivity of the lattice gauge theories with Wilson fermions.\cite{Menotti1987}
Although the timelike Wilson term (the term proportional to $r_\tau$) is artificial in the physics of a Weyl semimetal presented here, we cannot avoid this problem as far as we use the Wilson fermion formalism.

\subsection{The $\Pi_0$ instability}
In this case, we assume that $\langle N_n\rangle=-\sigma \bm{1}+\rho_0 \gamma_0\Pi_0$.
The procedure for the derivation of the free energy is the same as the case of $i\gamma_5$.
The mean-field decoupling of the interaction term is done to be
\begin{align}
\begin{split}
&e^{-\sum_n\mathrm{tr}\left[N_nP^+_0N_{n+\hat{0}}P^-_0\right]}\\
&\sim\exp\left\{-\sum_{n}\left[(1-r_\tau^2)\sigma^2+(1+r_\tau^2)\rho_0^2+\bar{\psi}_n\Gamma\psi_n\right]\right\},\label{int-term2}
\end{split}
\end{align}
where $\Gamma=\frac{1}{2}\left[(1-r_\tau^2)\sigma+(\gamma_0\Pi_0)^T(1+r_\tau^2)\rho_0\right]$.
Combining Eqs. (\ref{Effective-action}) and (\ref{int-term2}), the effective action expressed by the auxiliary fields $\sigma$ and $\rho_0$ is given by
\begin{align}
\begin{split}
S_{\mathrm{eff}}(\sigma,\rho_0)&={\textstyle \frac{V}{T}}\left[(1-r_\tau^2)\sigma^2+(1+r_\tau^2)\rho_0^2\right]\\
&\quad+\sum_k \bar{\psi}_k\mathcal{M}(\bm{k};\sigma,\rho_0)\psi_k, \label{S_eff_SCL2}
\end{split}
\end{align}
where the matrix $\mathcal{M}$ is given explicitly as
\begin{align}
\begin{split}
\mathcal{M}&=
\begin{bmatrix}
\tilde{m}(\bm{k})+r_\tau+b\sigma_3 & \sigma_j\sin k_j-\tilde{\rho_0}\\
-\sigma_j\sin k_j+\tilde{\rho_0} & \tilde{m}(\bm{k})+r_\tau-b\sigma_3
\end{bmatrix}\label{M2}
\end{split}
\end{align}
with $\tilde{\rho_0}=\frac{1}{2}(1+r_\tau^2)\rho_0$.
By calculating the determinant of $\mathcal{M}$, the free energy is obtained as
\begin{align}
\begin{split}
&\mathcal{F}(\sigma,\rho_0)=(1-r_\tau^2)\sigma^2+(1+r_\tau^2)\rho_0^2-\int_{-\pi}^{\pi}\frac{d^3k}{(2\pi)^3}\\
&\times\ln\left\{\frac{G(\bm{k},b,\tilde{\rho_0})G(\bm{k},-b,-\tilde{\rho_0})-H(\bm{k},b,\tilde{\rho_0})}{\left[\tilde{m}(\bm{k})+r_\tau\right]^2-b^2}\right\},\label{Feff-SCL2}
\end{split}
\end{align}
where $G(\bm{k},b,\tilde{\rho_0})=[\tilde{m}(\bm{k})+r_\tau-b]\times
\{[\tilde{m}(\bm{k})+r_\tau]^2-b^2+(\sin k_3-\tilde{\rho_0})^2\}
+s_\bot^2(\bm{k})[\tilde{m}(\bm{k})+r_\tau+b]$ and $H(\bm{k},b,\tilde{\rho_0})=4s_\bot^2(\bm{k})\{\tilde{\rho_0}[\tilde{m}(\bm{k})+r_\tau]+b\sin k_3\}^2$.
The ground state is determined by the stationary condition $\partial \mathcal{F}(\sigma,\rho_0)/\partial \sigma=\partial \mathcal{F}(\sigma,\rho_0)/\partial \rho_0=0$.

\subsection{The $\Pi_3$ instability}
In this case, we assume that $\langle N_n\rangle=-\sigma \bm{1}+\rho_3 \gamma_0\Pi_3$.
The procedure for the derivation of the free energy is the same as the case of $i\gamma_5$.
The mean-field decoupling of the interaction term is done to be
\begin{align}
\begin{split}
&e^{-\sum_n\mathrm{tr}\left[N_nP^+_0N_{n+\hat{0}}P^-_0\right]}\\
&\sim\exp\left\{-\sum_{n}\left[(1-r_\tau^2)\sigma^2+(1+r_\tau^2)\rho_3^2+\bar{\psi}_n\Gamma\psi_n\right]\right\},\label{int-term3}
\end{split}
\end{align}
where $\Gamma=\frac{1}{2}\left[(1-r_\tau^2)\sigma+(\gamma_0\Pi_3)^T(1+r_\tau^2)\rho_3\right]$.
Combining Eqs. (\ref{Effective-action}) and (\ref{int-term3}), the effective action expressed by the auxiliary fields $\sigma$ and $\rho_3$ is given by
\begin{align}
\begin{split}
S_{\mathrm{eff}}(\sigma,\rho_3)&={\textstyle \frac{V}{T}}\left[(1-r_\tau^2)\sigma^2+(1+r_\tau^2)\rho_3^2\right]\\
&\quad+\sum_k \bar{\psi}_k\mathcal{M}(\bm{k};\sigma,\rho_3)\psi_k, \label{S_eff_SCL3}
\end{split}
\end{align}
where the matrix $\mathcal{M}$ is given explicitly as
\begin{align}
\begin{split}
\mathcal{M}&=
\begin{bmatrix}
\tilde{m}(\bm{k})+r_\tau+b\sigma_3 & \sigma_j\sin k_j+i\tilde{\rho_3}\sigma_3\\
-\sigma_j\sin k_j+i\tilde{\rho_3}\sigma_3 & \tilde{m}(\bm{k})+r_\tau-b\sigma_3
\end{bmatrix}\label{M3}
\end{split}
\end{align}
with $\tilde{\rho_3}=\frac{1}{2}(1+r_\tau^2)\rho_3$.
By calculating the determinant of $\mathcal{M}$, the free energy is obtained as
\begin{align}
\begin{split}
&\mathcal{F}(\sigma,\rho_3)=(1-r_\tau^2)\sigma^2+(1+r_\tau^2)\rho_3^2-\int_{-\pi}^{\pi}\frac{d^3k}{(2\pi)^3}\\
&\times\ln\left\{\left[s^2(\bm{k})+\left[\tilde{m}(\bm{k})+r_\tau\right]^2-\mu\right]^2+4\mu s^2_\bot(\bm{k})\right\},\label{Feff-SCL3}
\end{split}
\end{align}
where $\mu=b^2-\tilde{\rho_3}^2$.
The ground state is determined by the stationary condition $\partial \mathcal{F}(\sigma,\rho_3)/\partial \sigma=\partial \mathcal{F}(\sigma,\rho_3)/\partial \rho_3=0$.

\subsection{The $\Pi_{1,2}$ instability}
In this case, we assume that $\langle N_n\rangle=-\sigma \bm{1}+\rho_1 \gamma_0\Pi_1$ or $\langle N_n\rangle=-\sigma \bm{1}+\rho_2 \gamma_0\Pi_2$ .
Since there exists a spin degree of freedom in the $xy$-plane, we show the calculation for $\Pi_1$.
The procedure for the derivation of the free energy is the same as the case of $i\gamma_5$.
The mean-field decoupling of the interaction term is done to be
\begin{align}
\begin{split}
&e^{-\sum_n\mathrm{tr}\left[N_nP^+_0N_{n+\hat{0}}P^-_0\right]}\\
&\sim\exp\left\{-\sum_{n}\left[(1-r_\tau^2)\sigma^2+(1+r_\tau^2)\rho_1^2+\bar{\psi}_n\Gamma\psi_n\right]\right\},\label{int-term4}
\end{split}
\end{align}
where $\Gamma=\frac{1}{2}\left[(1-r_\tau^2)\sigma+(\gamma_0\Pi_1)^T(1+r_\tau^2)\rho_1\right]$.
Combining Eqs. (\ref{Effective-action}) and (\ref{int-term4}), the effective action expressed by the auxiliary fields $\sigma$ and $\rho_0$ is given by
\begin{align}
\begin{split}
S_{\mathrm{eff}}(\sigma,\rho_1)&={\textstyle \frac{V}{T}}\left[(1-r_\tau^2)\sigma^2+(1+r_\tau^2)\rho_1^2\right]\\
&\quad+\sum_k \bar{\psi}_k\mathcal{M}(\bm{k};\sigma,\rho_1)\psi_k, \label{S_eff_SCL4}
\end{split}
\end{align}
where the matrix $\mathcal{M}$ is given explicitly as
\begin{align}
\begin{split}
\mathcal{M}&=
\begin{bmatrix}
\tilde{m}(\bm{k})+r_\tau+b\sigma_3 & \sigma_j\sin k_j+i\tilde{\rho_1}\sigma_1\\
-\sigma_j\sin k_j+i\tilde{\rho_1}\sigma_1 & \tilde{m}(\bm{k})+r_\tau-b\sigma_3
\end{bmatrix}\label{M4}
\end{split}
\end{align}
with $\tilde{\rho_1}=\frac{1}{2}(1+r_\tau^2)\rho_1$.
By calculating the determinant of $\mathcal{M}$, the free energy is obtained as
\begin{align}
\begin{split}
&\mathcal{F}(\sigma,\rho_1)=(1-r_\tau^2)\sigma^2+(1+r_\tau^2)\rho_1^2-\int_{-\pi}^{\pi}\frac{d^3k}{(2\pi)^3}\\
&\times\ln\left\{\frac{I(\bm{k},b,\tilde{\rho_1})I(\bm{k},-b,-\tilde{\rho_1})-J(\bm{k},b,\tilde{\rho_1})}{\left[\tilde{m}(\bm{k})+r_\tau\right]^2-b^2}\right\},\label{Feff-SCL4}
\end{split}
\end{align}
where $I(\bm{k},b,\tilde{\rho_1})=[\tilde{m}(\bm{k})+r_\tau-b]\times
\{[\tilde{m}(\bm{k})+r_\tau]^2+\sin^2k_3-b^2\}
+[\tilde{m}(\bm{k})+r_\tau+b]\times[\sin^2k_1+(\sin k_2+\tilde{\rho_1})^2]$ and
$J(\bm{k},b,\tilde{\rho_1})=4\sin^2k_3(\{\tilde{\rho_1}[\tilde{m}(\bm{k})+r_\tau]+b\sin k_2\}^2+b^2\sin^2k_1)$.
The ground state is determined by the stationary condition $\partial \mathcal{F}(\sigma,\rho_1)/\partial \sigma=\partial \mathcal{F}(\sigma,\rho_1)/\partial \rho_1=0$.

\section{Numerical Results \label{Sec-Results}}
\begin{figure}[!t]
\centering
\includegraphics[width=1.0\columnwidth,clip]{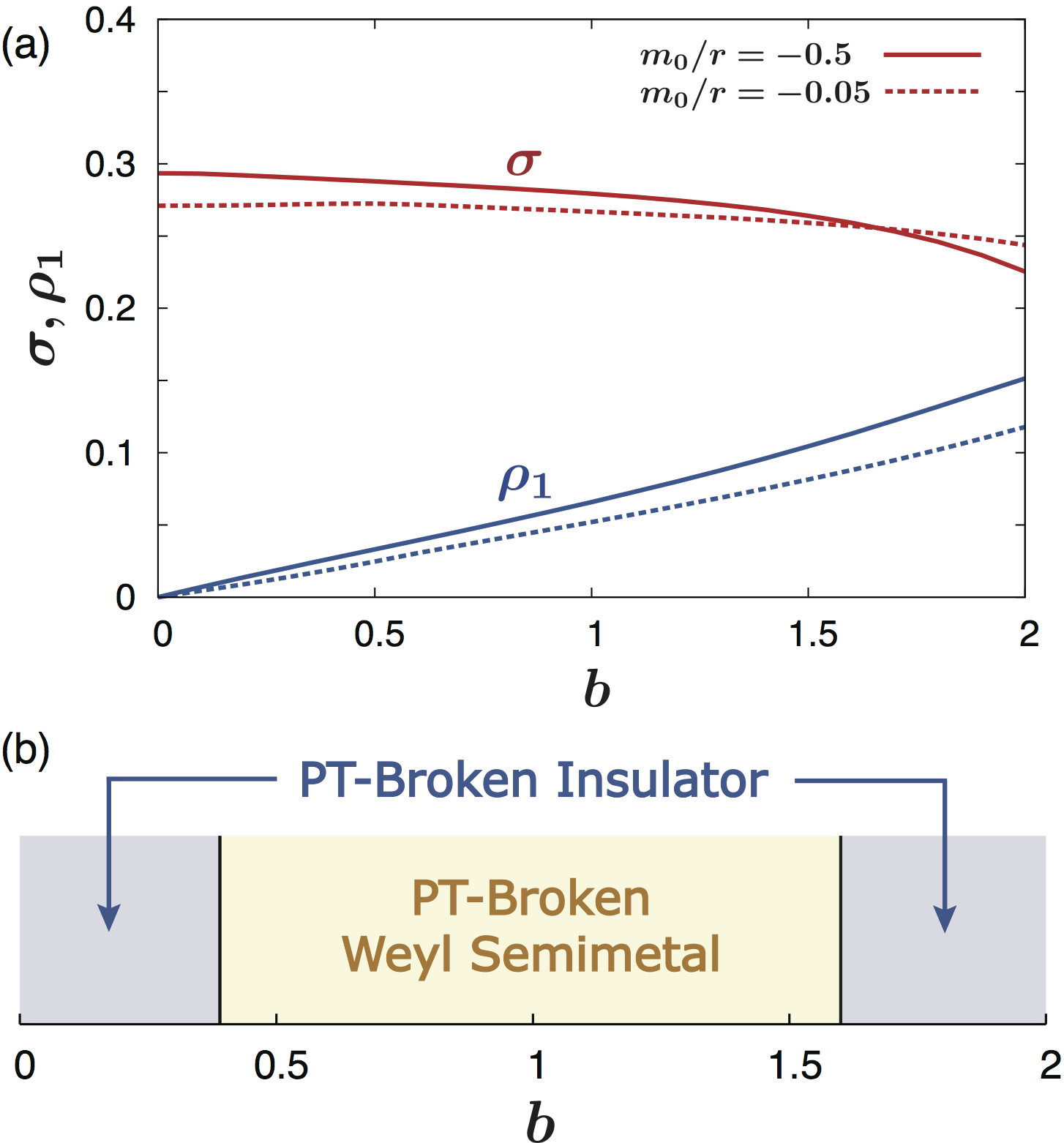}
\caption{(Color online) (a) $b$ dependences of $\sigma$ and $\rho_1$ with $r_\tau=0.5$ and $r=1$.
(b) Phase diagram in the strong coupling limit ($\beta=0$) with $m_0/r=-0.5$, $r_\tau=0.5$ and $r=1$.}\label{Fig1}
\end{figure}
For any set of $(m_0, b, r, r_\tau)$, it was found that the values of $\Delta$, $\rho_0$ and $\rho_3$ are always zero.
It was also found that the values of $\sigma$ and $\rho_1$ (or $\rho_2$)  are nonzero.
This means that the ground state of the system in the strong coupling limit is the parity and time-reversal symmetries (PT) broken phase signaled by nonzero $\rho_1$, and that the symmetry-broken phases signaled by nonzero $\Delta$, $\rho_0$ or $\rho_3$ do not arise.
Note that time-reversal symmetry of the system is originally broken in the non-interacting Weyl semimetal phase.

Throughout this paper, we set $r_\tau=0.5$ and $r=1$.
The $b$ dependence of $\sigma$ is shown in Fig. \ref{Fig1}(a).
$\sigma$ is a decreasing function of $b$.
The term proportional to $\sigma$ gives a correction to the bare mass $m_0$, i.e., is regarded as a mass renormalization induced by $1/r$ Coulomb interactions, as is seen from Eq. (\ref{m_k-tilde}).
Hence we should define the effective mass in the strong coupling limit as $m_{\rm eff}=m_0+\frac{1}{2}(1-r_\tau^2)\sigma$.
The renormalization becomes weaker as the time-reversal symmetry breaking perturbation $b$ becomes larger.
This is understood as follows: When the exchange coupling between the magnetic impurities and bulk electrons is strong, the bulk electrons prefer the ferromagnetic configuration.
Then the effective on-site interactions [the second term in Eq. (\ref{Effective-action})] become weaker and as a result, the renormalization effect becomes weaker.

The $b$ dependence of $\rho_1$ is also shown in Fig. \ref{Fig1}(a).
$\rho_1$ is an increasing function of $b$.
It is worthy to note that $\rho_1=0$ when $b=0$.
This means that time-reversal and parity symmetries of the system is preserved even in the strong coupling limit.
When $b=0$, we can distinguish by the $Z_2$ invariant whether the system is topologically nontrivial or trivial.
Namely, if $0>m_{\rm eff}/r>-2$ ($m_{\rm eff}/r>0$), then the system is topologically nontrivial (trivial).
In the case of $m_0/r=-0.5$, the effective mass is obtained as $m_{\rm eff}/r\simeq -0.39$.
This indicates that the topological insulator phase survives in the strong coupling limit.
On the other hand, in the case of $m_0/r=-0.05$, i.e. in the case of small $|m_0|$, the effective mass is obtained as $m_{\rm eff}/r\simeq 0.05$ and we see that the topological insulator phase changes to the normal insulator phase.

Let us look at the energy spectrum in the presence of the $\rho_1\Pi_1$ term, when $m_0/r=-0.5$.
In this case, the mean-field single-particle Hamiltonian is written as
\begin{align}
\begin{split}
\mathcal{H}(\bm{k})=\alpha_j\sin k_j+\tilde{m}(\bm{k})\alpha_4+b\Sigma_3+\rho_1\Pi_1.
\end{split}
\end{align}
We can obtain the energy spectrum analytically as
\begin{align}
\begin{split}
E(\bm{k})&=\pm\left\{s^2(\bm{k})+\left[\tilde{m}(\bm{k})\right]^2+b^2+\rho_1^2\right.\\
&\quad\left.\pm2\sqrt{\left[\tilde{m}(\bm{k})b-\rho_1\sin k_2\right]^2+(b^2+\rho_1^2)\sin^2k_3}\right\}^{1/2},\label{spectrum-PT-broken}
\end{split}
\end{align}
where $s^2(\bm{k})=\sum_{i=1}^3\sin^2k_i$.
By plotting this equation numerically, we find that there exists the region where the spectrum is gapless even when $\rho_1\neq 0$.
In the small $b$ region, there is a finite gap in the spectrum.
At $b\simeq 0.39$, the two bands of Eq. (\ref{spectrum-PT-broken}) start to touch at the momentum point $(k_1, k_2, k_3)\simeq(0, 0.005\pi, 0)$.
As $b$ is increased, the two Weyl nodes move from the $k_3=0$ point toward the $k_3=\pm\pi$ directions, crossing the $k_3$-axis.
The important point is that the two Weyl nodes do not exist at $k_2=0$ due to nonzero $\rho_1$.
Then at $b\simeq 1.60$, the two Weyl nodes meet at the point $(k_1, k_2, k_3)\simeq(0, -0.022\pi, \pi)$ and the band gap opens.
The phase diagram in the strong coupling limit is shown in Fig. \ref{Fig1}(b).
We call the gapped phases the parity and time-reversal symmetries (PT) broken insulator, and call the gapless phase the PT-broken Weyl semimetal.
In the PT-broken Weyl semimetal phase, we have confirmed that the $k_2$ dependence of the energy dispersions (\ref{spectrum-PT-broken}) near the band touching points is linear.

\section{Possible Global Phase Diagram \label{Sec-Phasediagram}}
Let us discuss a global phase diagram of a correlated Weyl semimetal.
First, we consider the phase diagram in the non-interacting limit ($\beta=\infty$) where the Hamiltonian is given by  Eq. (\ref{H0-lattice}).
When $b\neq 0$, time-reversal symmetry is broken and we cannot define the $Z_2$ invariant.
In this paper we call the phase with $0>m_0/r>-2$ the magnetic topological insulator.
This is because the phase transition from a topological insulator phase to another phase generally requires the gap closing.
Further, 3D topological insulator phases are characterized by the theta term with $\theta=\pi$.
A recent study shows that the value of $\theta$ remains $\pi$ even in the presence of the time-reversal symmetry breaking term $b\Sigma_3$ with not large $b$.\cite{Baasanjav2013}
Thus as far as the band gap opens, we call this phase the magnetic topological insulator.

As $b$ is increased, the energy bands touch and the two Weyl nodes start to split from the point $k_3=0$ toward $k_3=\pm\pi$ at $b=|m_0|$.
Then the Weyl nodes reach $k_3=\pm\pi$ to annihilate each other and the energy gap opens at $b=2r+m_0$.
These results are obtained from the equation for the appearance of the Weyl semimetal phase, $b^2=[m_0+r(1-\cos k_3)]^2+\sin^2 k_3$.
It is known that the Weyl semimetals have a nonzero Hall conductivity.
In the present case where the two Weyl nodes exist, as discussed in Ref. \citen{Burkov2011a}, the Hall conductivity of the system $\sigma^{\rm 3D}_{xy}$ is proportional to the distance of the two Weyl nodes in the momentum space $\Delta_{\rm W}$: $\sigma^{\rm 3D}_{xy}=e^2\Delta_{\rm W}/(2\pi h)$.
Thus in the gapped phase realized with $b>2r+m_0$, the Hall conductivity reaches $e^2/(ha)$ where $a$ is the lattice constant.
We call this phase the anomalous Hall insulator.

\begin{figure}[!t]
\centering
\includegraphics[width=1.0\columnwidth,clip]{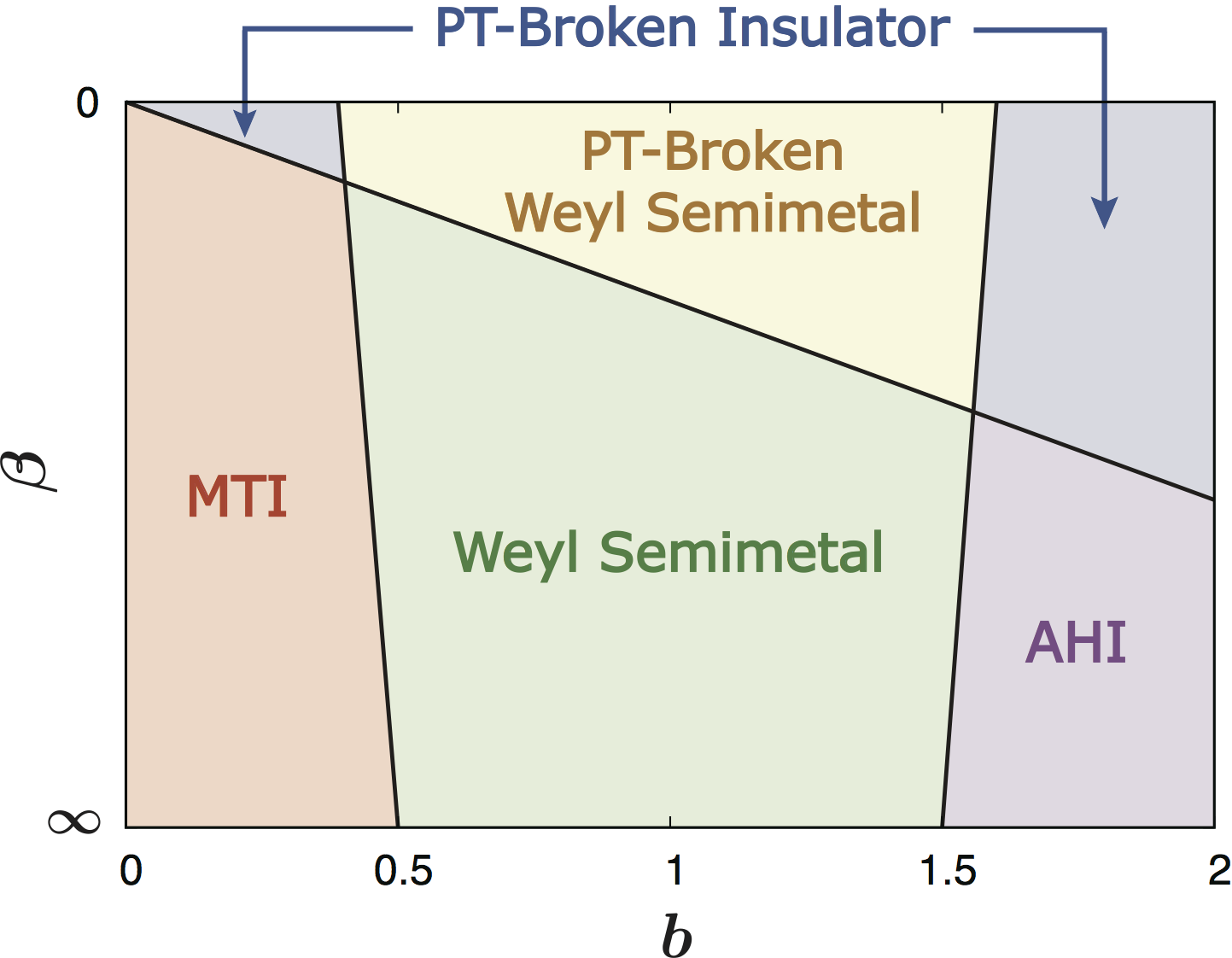}
\caption{(Color online) A possible global phase diagram of a correlated Weyl semimetal with $m_0/r=-0.5$, $r_\tau=0.5$ and $r=1$.
The phase boundary between the magnetic topological insulator (MTI) phase and the Weyl semimetal phase is determined by the condition $b=|m_{\rm eff}|$.
The phase boundary between the Weyl semimetal phase and the anomalous Hall insulator (AHI) phase is determined by the condition $b=2r+m_{\rm eff}$.
The parity and time-reversal symmetries (PT) broken Weyl semimetal phase and the PT-broken insulator phase are signaled by nonzero $\rho_1$( or $\rho_2$).
The phase boundaries between these two phases are determined (numerically) by the appearance or disappearance of the Weyl nodes.
The $\beta=0$ ($\beta=\infty$) line represents the strong coupling (non-interacting) limit.
}\label{Fig2}
\end{figure}

In the presence of $1/r$ Coulomb interactions, the mass $m_0$ is renormalized to be $m_{\rm eff}$.
This renormalization effect stabilizes the Weyl semimetal phase, i.e. makes the region of the Weyl semimetal phase larger.
When the interactions are strong enough, the order parameter $\rho_1$ (or $\rho_2$) for the PT-broken phase becomes nonzero and, as a result, the Weyl semimetal phase changes to another Weyl semimetal phase with broken parity and time-reversal symmetries.
From the numerical values in the strong coupling limit (see the previous section), the phase boundaries between the PT-broken Weyl semimetal and the PT-broken insulator phases seem to be approximated by $b\simeq |m_{\rm eff}|$ and $b\simeq 2r+m_{\rm eff}$.
It has been shown that the gapped phases (topological and normal insulators) with $b=0$ are stable in the strong coupling region.\cite{Sekine2013a}
Thus in the case of not small $|m_0|$ ($m_0<0$), the topological insulator phase would be continuous from the non-interacting limit to the strong coupling limit.
On the other hand, in the case of small $|m_0|$, the phase in the strong coupling limit is the normal insulator, although the phase in the non-interacting limit is the topological insulator.
In such a case, there would exist the coupling strength at which $m_{\rm eff}$ becomes zero.
By connecting the phase boundaries between the strong coupling region and the weak coupling region, we propose a possible global phase diagram of a correlated Weyl semimetal shown in Fig. \ref{Fig2}.

Here we would like to mention the relation between the proposed phase diagram and real materials.
As possible host materials, 3D topological insulators of Bi$_2$Se$_3$ family have large dielectric constant ($\epsilon \approx 100$)\cite{dielectric-constant}.
From this value, we estimate the strength of $1/r$ Coulomb interactions in Bi$_2$Se$_3$ family as $\beta\approx 3$.
This suggests that the Coulomb interactions in Bi$_2$Se$_3$-family-based Weyl semimetals are not strong.
Therefore it is difficult to predict the behavior of real materials from the present study.
However, the following could be mentioned.
When the value of $b$ is small, i.e. the exchange interactions with magnetic impurities are weak, parity symmetry of the system will not be broken.
On the other hand, when the value of $b$ is large, the PT-broken Weyl semimetal or the PT-broken insulator might be observed.
The energy spectra of Bi$_2$Se$_3$ family have been experimentally observed by angle-resolved photoemission spectroscopy (ARPES).
Thus those of Bi$_2$Se$_3$-family-based Weyl semimetals will be also observed when they are experimentally realized.
Theoretically, a Weyl semimetal phase in Bi$_2$(Se$_x$Te$_{1-x}$)$_3$ doped with magnetic impurities has been predicted.\cite{Kurebayashi2014}
As mentioned in the previous section, the PT-broken Weyl semimetal phase is characterized by the Weyl nodes located in the points which are deviated from the original location.
It is expected that such a locational deviation of Weyl nodes can be observed by ARPES.

\section{Discussions \label{Sec-Discussion}}
Let us consider the possible gapped phases in our model.
From our numerical results and the discussion in Ref. \citen{Burkov2011}, we see that only the $\alpha_5$ instability can lead to a gapped phase among the five instabilities $\alpha_5$ and $\Pi_\mu$.
In the presence of the $\alpha_5$ term, the mean-field Hamiltonian is written as
\begin{align}
\begin{split}
\mathcal{H}(\bm{k})=\alpha_j\sin k_j+\tilde{m}(\bm{k})\alpha_4+b\Sigma_3+\Delta\alpha_5,
\end{split}
\end{align}
and we can obtain the energy spectrum analytically as
\begin{align}
\begin{split}
E(\bm{k})&=\pm\sqrt{s^2_\bot(\bm{k})+\left[\sqrt{[\tilde{m}(\bm{k})]^2+\Delta^2+\sin^2 k_3}\pm b\right]^2},\label{spectrum-alpha5}
\end{split}
\end{align}
where $s^2_\bot(\bm{k})=\sum_{i=1}^2\sin^2k_i$.
In this case, the two Weyl nodes arise on the $k_3$ axis when the conditions $k_1=k_2=0$ and $b^2>[\tilde{m}(0,0,k_3)]^2+\Delta^2$ are satisfied.
Conversely, the energy gap opens when $\Delta^2>b^2-[\tilde{m}(0,0,k_3)]^2$ is satisfied.
Actually, such a gapped phase has been obtained in a Weyl semimetal with short-range interactions.\cite{Sekine2013}

Further, we note that the inter-nodal scattering, which can lead to gap openings,\cite{Yang2011,Wei2012} is contained in the low-energy limit of our model.
In the following, we briefly discuss this process in our model.
We express $\psi_k$ in terms of the annihilation operator in the $\lambda$-th band $a_{k\lambda}$ and the eigenfunction of the $\lambda$-th band $|u_{k\lambda}\rangle$ as $\psi_{k}=\sum_{\lambda=1}^4 a_{k\lambda}|u_{k\lambda}\rangle$.
Here we label the two bands near the Fermi level as $\lambda=2,3$.
Then we can approximate $\psi_n$ in the low-energy limit as
\begin{align}
\begin{split}
\psi_n&\simeq \left(\int_{|\bm{k}-W_+|<\Lambda}\frac{d^3k}{(2\pi)^3}+\int_{|\bm{k}-W_-|<\Lambda}\frac{d^3k}{(2\pi)^3}\right)e^{i\bm{k}\cdot\bm{r}_n}\sum_{\lambda=2,3} a_{k\lambda}|u_{k\lambda}\rangle\\
&\equiv e^{iQz}\psi_{R,n}+e^{-iQz}\psi_{L,n},
\end{split}
\end{align}
where $\Lambda$ is a momentum cutoff, $W_\pm=(0,0,\pm Q)$ with $Q=\sqrt{b^2-[\tilde{m}(0,0,k_3)]^2}$, and $\psi_{R(L),n}$ is the annihilation operator around the Weyl node $W_{+(-)}$.
With the use of this expression, the mean-field decoupled interaction term $\bar{\psi}_n\langle N_n\rangle\psi_n$ is written as
\begin{align}
\begin{split}
\bar{\psi}_n\langle N_n\rangle\psi_n&\simeq \bar{\psi}_{R,n}\langle \bar{\psi}_{R,n}\psi_{R,n}\rangle\psi_{R,n}+\bar{\psi}_{R,n}\langle \bar{\psi}_{L,n}\psi_{R,n}\rangle\psi_{L,n}\\
&\quad +\bar{\psi}_{R,n}\langle \bar{\psi}_{L,n}\psi_{L,n}\rangle\psi_{R,n}+\bar{\psi}_{L,n}\langle \bar{\psi}_{R,n}\psi_{R,n}\rangle\psi_{L,n}\\
&\quad +\bar{\psi}_{L,n}\langle \bar{\psi}_{R,n}\psi_{L,n}\rangle\psi_{R,n}+\bar{\psi}_{L,n}\langle \bar{\psi}_{L,n}\psi_{L,n}\rangle\psi_{L,n},
\end{split}
\end{align}
where we have omitted the oscillating terms.
The terms $\bar{\psi}_R\langle\bar{\psi}_L\psi_R\rangle\psi_L$ and $\bar{\psi}_L\langle\bar{\psi}_R\psi_L\rangle\psi_R$ can be a mass term in the low-energy effective model.
If the mass term is induced by interactions in the low-energy effective model, then the energy gap opens and the Weyl semimetal phase is broken.
Therefore, note that the effects which can lead to the gapped phase are taken into account in our calculation of a bulk model for a correlated Weyl semimetal.
However, our result suggests the existence of the gapless phase, the PT-broken Weyl semimetal phase in the strong coupling limit.

Finally, we mention the difference between the Weyl semimetal and the graphene.
In graphene, there are also gapless linear dispersions which can be described by the Weyl Hamiltonian around two inequivalent momentum points.
It is known that parity symmetry breaking in graphene leads to a gap opening.\cite{Semenoff1984,Haldane1988,Fuchs2007}
However, in the case of the Weyl semimetal, the system can remain gapless even when parity symmetry is (spontaneously) broken.
This results from the topological nature of the Weyl semimetal, namely that a gap opens only when the Weyl nodes with opposite chirality meet each other.

\section{Summary \label{Sec-Summary}}
To summarize, based on the U(1) lattice gauge theory, we have studied the effects of strong $1/r$ long-range Coulomb interactions in a Weyl semimetal with broken time-reversal symmetry.
We have considered all the possible 16 instabilities within the mean-field approximation.
It was shown that parity symmetry of the system is spontaneously broken but the Weyl semimetal phase, which is different from the non-interacting phase, survives in the strong coupling limit.
We have presented a possible global phase diagram of a correlated Weyl semimetal.
From the proposed global phase diagram, it is expected that the Weyl semimetal phase is stabilized by $1/r$ long-range Coulomb interactions.
In this study, the number of the Weyl nodes is two.
It would be interesting to study the correlation effects in Weyl semimetals with more than two nodes.

\section*{Acknowledgments}
\begin{acknowledgments}
A.S. is supported by the JSPS Research Fellowship for Young Scientists.
This work was supported by Grant-in-Aid for Scientific Research (No. 25103703, No. 26107505 and No. 26400308) from the Ministry of Education, Culture, Sports, Science and Technology (MEXT), Japan.
\end{acknowledgments}


\end{document}